\def \vc #1{{\mbox{\boldmath $#1$}}}

\def\thetag{{\vc \theta}}

\def\deltag{{\vc \delta}}
\def\gammag{{\vc \gamma}}

\documentclass[runningheads]{svmult}

\usepackage{makeidx}   
\usepackage{graphicx}  
\usepackage{subeqnar}  
\usepackage{multicol}  
\usepackage{physprbb}  
\makeindex             

\begin{document}
\title*{Cosmological weak lensing}
\toctitle{Focusing of a Parallel Beam to Form a Point
\protect\newline in the Particle Deflection Plane}
%
%
\titlerunning{Focusing of a Parallel Beam}
%
\author{Yannick Mellier\inst{1,2}
\and Ludovic van Waerbeke\inst{1,3}
\and Francis Bernardeau\inst{4}
\and Ismael Tereno\inst{1,5}}
\authorrunning{Y. Mellier et al.}
%
%
\institute{IAP, 98bis boulevard Arago, 75014 Paris, France
\and Observatoire de Paris, LERMA,
     61 avenue de l'Observatoire,
     75014 Paris, France
\and CITA, 60 St George Str., M5S 3H8, Toronto, Canada,
\and SPhT, CE Saclay, 91191 Gif-sur-Yvette Cedex,
\and Department of Physics, University of Lisbon
1749-016 Lisboa, Portugal.}

\maketitle              

\begin{abstract}
We present the current status of cosmic shear studies and their 
  implications on cosmological models.
  Theoretical expectations and observational results are discussed
  in the framework of standard cosmology and CDM scenarios.  
   The potentials of the
next generation cosmic shear surveys are discussed.
\end{abstract}

\section{Introduction}
The gravitational deflection of light beams by 
large scale structures of the universe (cosmological lensing) 
  amplifies and modifies 
 the shape of distant galaxies and quasars. Magnification produces 
  correlation between the density of foreground lenses and  the 
  apparent luminosity of distant galaxies or quasars 
  (magnification bias), whereas distortion 
  induces a correlation of ellipticity distribution of lensed galaxies
(cosmic shear). In both cases, the properties of cosmological lensing signals
   probe the matter content  and the geometry of universe
  and how perturbations grew and clustered during the past
Gigayears.
\\
Albeit difficult to detect, the recent cosmic shear detections 
   claimed by several groups demonstrate that it is no longer 
  a technical challenge. It is therefore possible to study the
universe through a new  
   window which directly probes dark matter instead of light and allows
cosmologists 
 to measure  cosmological parameters 
  and dark matter power spectrum from weak gravitational distortion.

\section{Weak cosmological lensing}
Let us assume that the shape of galaxies can be simply characterize
  by their surface brightness second moments $I\left(\thetag\right)$,
  (see \cite{mel99}, \cite{bart01} and references therein):
\begin{equation}
M_{ij}= \displaystyle{ {\int
I\left(\thetag\right) \theta_i \ \theta_j
 \ {\rm d}^2\theta \over \int I(\thetag) \ {\rm d}^2\theta }} \ . 
\end{equation}
Because of gravitational lensing, 
  a galaxy with intrinsic ellipticity ${\bf e}$ is measured with
an ellipticity ${\bf e}+{\deltag}$, where ${\deltag}$ is the 
  gravitational distortion,
\begin{equation}\deltag =2 \gammag \ {(1-\kappa)
\over
(1-\kappa)^2+\vert \gammag \vert^2} \ =\left(\delta_1:{M_{11}-M_{22}
\over Tr(M)} \ ;  \ \delta_2:{ 2 M_{12} \over Tr(
M)} \right) \ .
\label{distor}
\end{equation}
$\kappa$ and $\gammag$ are respectively the gravitational convergence
 and shear. Both depend on the second derivatives of the
projected gravitational potential, $\varphi$:
\begin{equation}
 \kappa(\thetag) = \frac{1}{2}\,(\varphi_{,11}+\varphi_{,22})  ; \ \ \
 \gamma_1(\thetag) =
  \frac{1}{2}\left(\varphi_{,11}-\varphi_{,22}\right) ; \ \ \
  \gamma_2(\thetag) = \varphi_{,12} \ .
\label{convshear}
\end{equation}
   In the case of weak
lensing, $\kappa<<1$,  $\vert \gamma \vert <<1$ and $\delta \approx 2 \
\gamma$. Since large-scale structures have very low density 
 contrast, this linear relation is in particular valid on cosmological scales.
\\
Light propagation through an inhomogeneous universe accumulates
 weak lensing effects over Gigaparsec distances. 
 Assuming structures formed from gravitational growth of
    Gaussian fluctuations, cosmological weak lensing can be predicted
from Perturbation
  Theory at large scale.  To first order, the convergence $\kappa(\thetag)$ at
angular
position $\thetag$ is given by the line-of-sight integral
\begin{equation}
\kappa(\thetag)={3 \over 2} \Omega_0  \int_0^{z_s} n(z_s)  {\rm d}z_s
\int_0^{\chi(s)}  {D\left(z,z_s\right) D\left(z\right) \over
D\left(z_s\right)}  \delta\left(\chi,\thetag\right) \
\left[1+z\left(\chi\right)\right]  {\rm d}\chi 
\end{equation}
where $\chi(z)$ is the radial distance out to redshift $z$, $D$ the
angular diameter
distances, $n(z_s)$ is the redshift
distribution of the sources.
 \ \ \   $\delta$ is 
the mass density contrast responsible for the deflection at redshift
 $z$. Its amplitude at a given redshift  depends on the properties  of the
 power spectrum and its evolution with look-back-time.
\\
The cumulative weak lensing effects of
structures induce a shear field  
  which is primarily related to the   power spectrum of the projected
mass density, $P_\kappa$. 
   Its statistical properties can be recovered by
    the
shear top-hat variance \cite{me91,b91,k92}, 
\begin{equation}
\langle\gamma^2\rangle={2\over \pi\theta_c^2} \int_0^\infty~{{\rm
d}k\over k} P_
\kappa(k)
[J_1(k\theta_c)]^2,
\label{theovariance}
\end{equation}
the aperture mass
variance \cite{k94,sch98}
\begin{equation}
\langle M_{\rm ap}^2\rangle={288\over \pi\theta_c^4} \int_0^\infty~{{\rm
d}k\over k^3}
P_\kappa(k) [J_4(k\theta_c)]^2,
\label{theomap}
\end{equation}
 and the shear correlation
function \cite{me91,b91,k92}:

%
\begin{equation}
\langle\gamma\gamma\rangle_\theta={1\over 2\pi} \int_0^\infty~{\rm d} k~
k P_\kappa(k) J_0(k\theta),
\label{theogg}
\end{equation}
where $J_n$ is the Bessel function of the first kind.
 Higher order statistics, like  the skewness
of the convergence, $s_3(\theta)$, can also be computed. 
  They probe non Gaussian
features in the projected mass density field, like 
    massive clusters or compact groups of galaxies.
 (see \cite{bern97};
\cite{jain97} and references therein).  The amplitude 
 of cosmic shear signal and its sensitivity to cosmology
  can be  illustrated in the fiducial case of a power
law mass power spectrum with no cosmological constant and
a background population at a single redshift
$z$. In that case  $<\kappa(\theta)^2>$ and $s_3(\theta)$ write:
\begin{equation}
\label{eqvar}
<\kappa(\theta)^2>^{1/2} =<\gamma(\theta)^2>^{1/2} \approx 1\% \
 \sigma_8 \ \Omega_m^{0.75} \
z_s^{0.8
} \left({\theta \over  1'}\right)^{-\left({n+2 \over 2}\right)}  \ ,
 \
\end{equation}
and
\begin{equation}
\label{eqs3}
s_3(\theta)={\langle\kappa^3\rangle\over \langle\kappa^2\rangle^2}
\approx 40 \
  \Omega_m^{-0.8} \ z_s^{-1.35}  \ ,
\end{equation}
 where $n$ is the spectral index of
the power
spectrum of density fluctuations. Therefore, in principle the
      degeneracy between $\Omega_m$ and $\sigma_8$ can be  broken
   when both the variance and the skewness of the convergence
  are measured. 
\section{Detection of weak distortion signal}
\subsection{Observational challenge}
Eq. (\ref{eqvar}) shows that  the amplitude of weak lensing signal is
of the order of few percents, which is 
   much smaller than the
  intrinsic dispersion of ellipticity distribution of galaxies.
 van Waerbeke et al (\cite{vwal99}) explored which  strategy would be
best suited to probe statistical properties  of such a small signal.
  They have shown  that the   variance of $\kappa$ 
 can be measured with a survey covering about 1 $deg^2$,
  whereas for the skewness one needs at least 10 $deg^2$. Furthermore,
   more than 100 $deg^2$ must be observed in order to 
   uncover  information on $\Omega_{\Lambda}$
  or  the shape
  of the power spectrum over scales larger than 1 degree.
  For $\Omega_m=0.3$ and $\sigma_8=1$, the limiting shear amplitude
 can be simply expressed as follows
\begin{equation}
\label{survey}
<\gamma(\theta)^2>^{1/2} = 0.3\%  \ \left[{A_T \over 100
\ deg^2}\right]^{{1 \over 4}} \times \left[{\sigma_{\epsilon_{gal}}
\over 0.4} \right] \times
 \left[{n \over 20}\right]^{-{1 \over 2}} \times
 \left[{\theta \over 10'}\right]^{{-{1 \over 2}}} \ ,
\end{equation}
where $A_T$ is the total sky coverage of the
 survey. The numbers given in the brackets correspond to a measurement
at $3-\sigma$ confidence level of the shear variance. Eq. (\ref{survey})
  contains the specifications of a cosmic shear survey.
\subsection{First detection of weak distortion}
\begin{figure}[t]
\begin{center}
\includegraphics[width=9cm,angle=270]{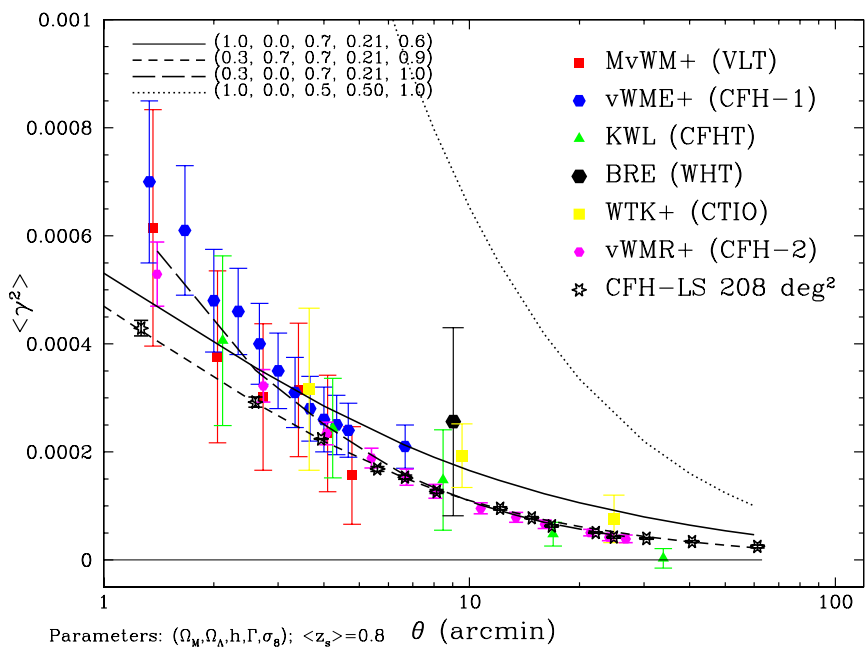}
\end{center}
\caption[]{Top hat variance of shear as function of angular scale from
  6 cosmic shear surveys. The CFHT-LS
(open black stars) illustrates
 the expected signal from a large survey covering 200 deg$^2$. For
most
 points the errors are smaller than the stars.
}
\label{sheartop}
\end{figure}

Despite technical limitations discussed above, 
 on scale significantly smaller than one degree, 
non-linear structures dominate and increase the amplitude of the lensing
signal, making its measurement easier.
  Few teams  started such surveys
  during the past years and succeeded to get a significant signal.
 Table \ref{tabcs} lists some published results. 
  Since each group used different telescopes and 
  adopted different observing strategy and 
  data analysis techniques,  one can figure out 
  the reliability of the final results.
\begin{table}
\begin{center}
{\small
\caption{Present status of cosmic shear surveys with published
results.
}
\label{tabcs}
\begin{tabular}{lcccl}\hline
Telescope& Pointings & Total Area & Lim. Mag. & Ref.. \\
\hline
CFHT & 5 $\times$ 30' $\times$30'& 1.7 deg$^2$ & I=24. &
\cite{vwal00}[vWME+]\\
CTIO & 3 $\times$ 40' $\times$40'& 1.5 deg$^2$ & R=26. &
\cite{wit00a}[WTK+]\\
WHT & 14 $\times$ 8' $\times$15'& 0.5 deg$^2$ & R=24. &
\cite{bacon00}[BRE]\\
CFHT & 6 $\times$ 30' $\times$30'& 1.0 deg$^2$ & I=24. &
\cite{kais00}[KWL]\\
VLT/UT1 & 50 $\times$ 7' $\times$7'& 0.6 deg$^2$ & I=24. &
\cite{mao01}[MvWM+]\\
HST/WFPC2 & 1 $\times$ 4' $\times$42'& 0.05 deg$^2$ & I=27. &
\cite{rhodes01}\\
CFHT & 4 $\times$ 120' $\times$120'& 6.5 deg$^2$ & I=24.
&\cite{vwal01}[vWMR+]\\
HST/STIS & 121 $\times$ 1' $\times$1'& 0.05 deg$^2$ & V$\approx 26$
& \cite{hammerle01}\\
CFHT & 5 $\times$ 126' $\times$140'& 24. deg$^2$ & R=23.5& \cite{hoek01a}
\\
CFHT & 10 $\times$ 126' $\times$140'& 53. deg$^2$ & R=23.5& \cite{hoekstra02}
\\
CFHT & 4 $\times$ 120' $\times$120'& 8.5 deg$^2$ & I=24.&
\cite{pen01}\\
HST/WFPC2 & 271 $\times$ 2.1 $\times$ 2.1 & 0.36 deg$^2$ & I=23.5 & \cite{ref02} \\
Keck+WHT & 173 $\times$ 2' $\times$ 8' & 1.6 deg$^2$ & R=25 & \cite{bacon02} \\ 
 & +13 $\times$ 16' $\times$ 8' &  &  &  \\ 
 & 7 $\times$ 16' $\times$ 16' & &  &  \\ 
\hline
\end{tabular}
}
\end{center}
\end{table}
Figure \ref{sheartop} show that all these 
  independent results are in very good
agreement\footnote{ \cite{hoek01a} data are missing because depth is
  different so the
sources are at lower redshift and the amplitude of the shear
  is not directly comparable to other data plotted}.  This  is 
  a convincing demonstration that the 
  expected correlation of ellipticities is real.
\subsection{Nature of the weak distortion signal}
The detection of coherent signal is not a demonstration of its
very nature. Even if a cosmological signal were expected, it could be
 contaminated by systematics, like  optical and atmospheric distortions,
 which mix together with the gravitational shear.  
 Contrary to  lensing effects, these contributions 
   are visible also on stars and can be corrected (using for 
  example the KSB method, \cite{ksb}). 
     However, 
  stars often show strong anisotropic shape  with elongation
 much larger than the expected amplitude of  the gravitational
distortion.
 The reliability of artificial anisotropy corrections is therefore 
  a critical step of the weak lensing analysis
 (see for example \cite{erben01}, \cite{vwal00} \cite{bacon00} and
\cite{bacon01}).  An elegant way 
   to check whether corrections are correctly done and to 
   confirm the   gravitational nature of the signal is 
  to decompose the signal into  E- and B- modes.
   The E-mode 
 contains signal produced by gravity-induced distortion whereas 
the B-mode is a pure curl-component, so it  
  only contains intrinsic ellipticity correlation or 
systematics residuals. Both modes have been extracted using the 
aperture mass statistics by van Waerbeke et al (\cite{vwal01},
 \cite{vw02}) and 
 Pen et al (\cite{pen01}) in the VIRMOS-DESCART survey as well as 
  by Hoekstra et al (\cite{hoekstra02}) in the Red Cluster Sequence survey.
 In both samples, the E-mode dominates the signal, although a small
  residual is detected in the B-mode. This strongly supports the 
   gravitational origin of the distortion.
\\
An alternative to gravitational lensing effect could be an 
   intrinsic correlations of ellipticities of galaxies produced 
  by proximity effects. It
  could results from galaxy formation processes.
   Several
  recent  numerical and theoretical  studies (see for
 example \cite{crit00}; \cite{mackey01})
 show that intrinsic correlations are negligible on scales beyond
  one arc-minute, provided that the survey is deep enough. In that
case,
  most lensed galaxies along a line of sight are spread over
 Gigaparsec scales and have no physical relation with its apparent
  neighbors.
  Hence, since most cosmic survey are deep, they are almost
    free of intrinsic correlations.
 We therefore are confident that the signal measured by all teams 
  is a genuine cosmic shear signal.
\section{Cosmological interpretations}
\subsection{2-point statistics and variance}
A comparison of the top-hat variance of shear 
   with some realistic cosmological models
 is ploted in Figure \ref{sheartop}. The amplitude of the shear
  has been scaled using photo-$z$ which gives $<z> \approx 1$.
 On this plot, we see that
 standard COBE-normalized CDM is ruled  at a 10$-\sigma$  confidence
level.  However,
  the degeneracy between $\Omega_m$ and
   $\sigma_8$  discussed in the previous section still hampers
  a strong discrimination among most popular cosmological models.
    The present-day constraints resulting from independent analyses
  by Maoli et al (\cite{mao01}),  Rhodes et al (\cite{rhodes01}), 
  van Waerbeke et 
 al (\cite{vwal01,vw02}), Hoekstra et al (\cite{hoek01a,hoekstra02}) 
  and R\'efr\'egier et al (\cite{ref02})  can be summarized by the following 
  conservative  boundaries (90\% confidence level):
\begin{equation}
0.05 \le \Omega_m \le 0.8 \ \ \ \ {\rm and} \ \ \ \ 0.5 \le \sigma_8
\le 1.2 \ ,
\end{equation}
 and, in the case of a flat-universe with $\Omega_m=0.3$,  they 
  lead to 
  $\sigma_8 \approx 0.9$.
\\
\begin{figure}[t]
\begin{center}
\includegraphics[width=9cm]{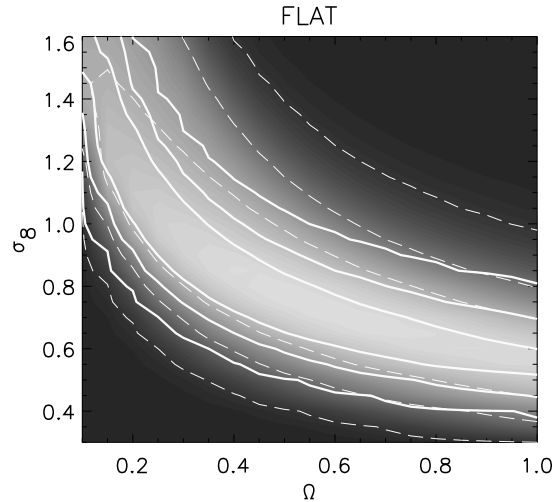}
\end{center}
\caption[]{Constraints on
  $\Omega$ and $\sigma_8$ for the flat cosmologies.  The confidence
  levels are $[68,95,99.9]$ from the brightest to the darkest area.
  The gray area and
  the dashed contours correspond to the computations with a full
  marginalisation over the default prior $\Gamma\in [0.05,0.7]$ and
  $z_s\in [0.24,0.64]$. The thick solid line contours are obtained from
 the prior $\Gamma\in [0.1,0.4]$ and $z_s\in [0.39,0.54]$ (which is a
  mean redshift $\bar z_s\in [0.8,1.1]$). From
  van Waerbeke et al. (2002).
  }
\label{omegasig}
\end{figure}
\begin{figure}[t]
\begin{center}
\includegraphics[width=10cm]{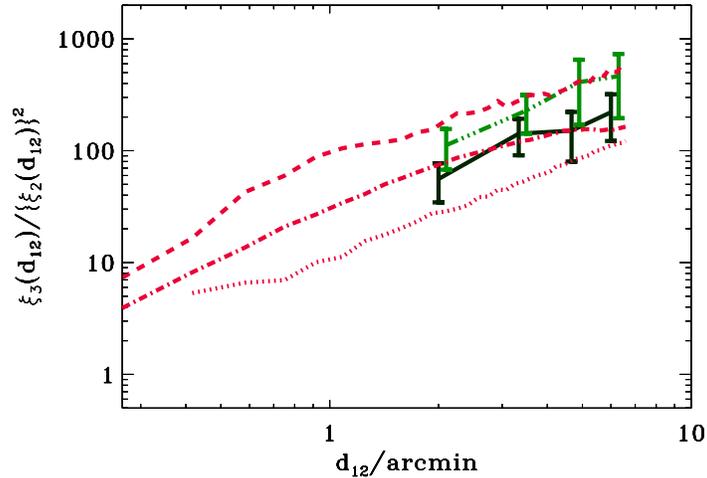}
\end{center}
\caption[]{Results for the VIRMOS-DESCART survey of the reduced three
point correlation function (\cite{bern02b}). The solid line
 with error bars shows the raw results, when both the $E$ and $B$
contributions
to the two-point correlation functions are included.  The  dot-dashed
 line with error bars corresponds to measurements where the contribution
of the $B$
mode has been subtracted out from the two-point correlation function.
 These
measurements are compared to results obtained in $\tau $CDM, OCDM and
$\Lambda$CDM simulations (dashed, dotted and dot-dashed lines
respectively).  }
\label{xi3}
\end{figure}
\subsection{The 3-point shear correlations function: breaking the
$\Omega_{{\rm m}}-\sigma_8$ degeneracy}
The measurement of non-Gaussian features needs informations
  on higher order statistics than variance. Although
  the afore mentioned  skewness of $\kappa$
   looks a promising quantity for this purpose, its measurements 
   suffers from a number a practical difficulties which are not yet 
   fixed.    Recently,
Bernardeau, van Waerbeke \& Mellier (\cite{bern02a})  have
proposed an alternative method using some specific patterns in the shear
three-point function.
 Despite the complicated shape of the
three-point correlation pattern, they uncovered 
 it can be used for the measurement
of non-Gaussian
 features. Their  detection strategy based on their method has
been tested on ray tracing simulations and 
  turns out  to be robust, usable in patchy catalogs, and quite
insensitive to the topology of the survey.  
\\
Bernardeau, Mellier \& van Waerbeke (\cite{bern02b}) used the analysis 
  of the  3-point correlations function on the VIRMOS-DESCART 
data.  Their results on  Figure \ref{xi3} show a 
2.4$\sigma$ signal over four independent angular bins, or
equivalently, a 4.9-$\sigma$ confidence level detection with respect
to measurements errors on scale of about $2$ to $4$ arc-minutes.
The amplitude and the shape of the signal are consistent
with theoretical expectations obtained from ray-tracing simulations.
This result supports
the idea that the measure corresponds to a  cosmological signal due to
the gravitational instability dynamics. Moreover, 
  its properties could be used
to put constraints on the cosmological parameters, in particular on
the density parameter of the Universe. Although the 
   errors 
 are  still large to permit secure conclusions, one clearly see that 
   the amplitude and the shape of the 3-point correlations function 
   match the most likely cosmological models.  Remarkably, the 
  $\Lambda$CDM scenario perfectly fit the data points.
\\
The Bernardeau et al.  (\cite{bern02b}) result is the 
first detection of non-Gaussian features in a cosmic shear survey and 
  it opens the route to break the $\Omega_{{\rm m}}-\sigma_8$ degeneracy.
 Furthermore, this method is weakly dependent on other parameters, 
  like the cosmological constant or the properties of the power
spectrum.   
 However, there are still some caveats which may be considered
seriously.
 One difficulty is the source  clustering which could 
   significantly perturb high-order statistics
 (Hamana et al 2000, \cite{hamana00}).
 If so, multi-lens plane cosmic shear analysis will be necessary
  which implies a good knowledge of the redshift distribution.
   For very deep cosmic shear surveys, this could be  could be
  a challenging issue.  
\section{What next?}
\begin{figure}[t]
\begin{center}
\includegraphics[width=7.5cm]{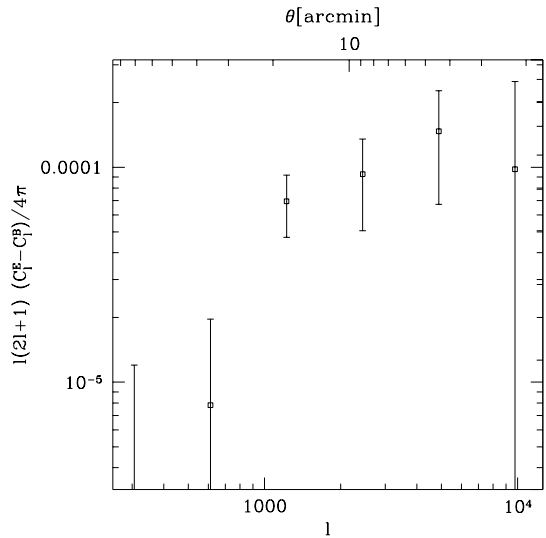}
\end{center}
\caption[]{Cosmological results from cosmic shear surveys:
 The angular power spectrum of the convergence field
  from the VIRMOS-DESCART  survey is plotted (From Pen et al 2001). 
  These are the first $C(l)$ of dark matter ever measured in cosmology.}
\label{cl}
\end{figure}

\begin{figure}[t]
\begin{center}
\includegraphics[width=8cm]{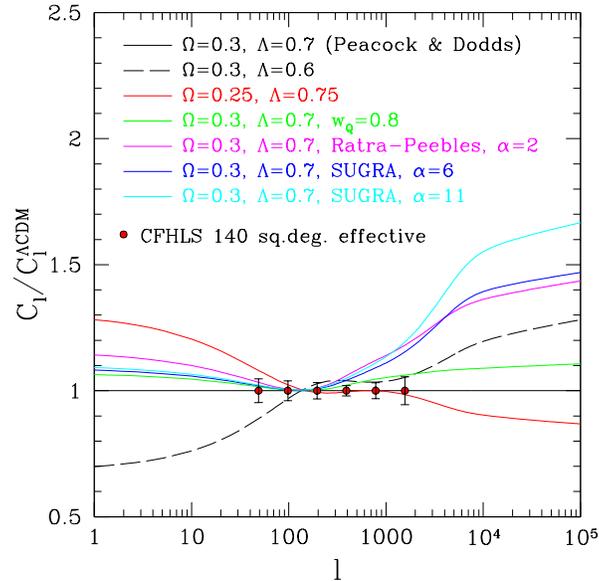}
\end{center}
\caption[]{The future of from cosmic shear surveys:
  Theoretical expectations of the CFHT Legacy Survey. The dots with
error
bars are the expected measurements of cosmic shear data from the 208
deg$^2$ of the CFHT Legacy Survey. The lines shows various models 
  discussed by Benabed \& Bernardeau (\cite{benab01}).}
\label{future1}
\end{figure}

 Because on going surveys increase 
  both in solid angle and in number of galaxies, they 
  will quickly  improve the accuracy of cosmic shear measurements, at
 a level where $\Omega_m$ and $\sigma_8$ will be known with  a 
  10\% accuracy.   Since it is based on gravitational deflection 
  by intervening matter spread over cosmological scales, the shape of 
   the distortion field also probes directly the shape of the 
  projected power spectrum of the (dark) matter. Pen et al (\cite{pen01})
already explored its properties measuring for the first 
  time the $C(l)$ of the dark matter (see Figure 
 \ref{cl}). We therefore know this is feasible with present data. 
However, we expect much more within the next decade. Surveys 
  covering  
   hundreds of degrees, with multi-bands data in order
to get redshift of sources and possibly detailed information of
 their clustering properties, are scheduled.  The CFHT Legacy
Survey\footnote{http://www.cfht.hawaii.edu/Science/CFHLS/} will cover
  200 deg$^2$ and is one of those
next-generation cosmic shear survey. Figures 1 and \ref{future1} shows
it 
  potential for cosmology. On figure 1 we simulated the expected signal
to noise of the shear variance as function of angular scale for 
  a $\Lambda$CDM cosmology.  The error bars are considerably reduced as
compared
 to present-day survey.  On Fig. \ref{future1}, we compare the expected
signal to noise 
 ratio of the CFHT Legacy Survey with the expected amplitude of the
angular 
power spectrum for several theoretical quintessence fields models. It
shows 
that even with 200 deg$^2$ which include multi-color informations in
order to
get redshift of sources, one can already obtain interesting constraints
on 
  cosmology beyond standard models.\\
The use of cosmic shear data can be much more efficient if they are used
together
  with other surveys, like CMB (Boomerang, MAP, Planck), SNIa surveys,
or even 
galaxy surveys (2dF, SDSS).  For example, SDSS will soon provide 
  the 100, 000 quasars with redshifts.  M\'enard \& Bartelmann 
 (\cite{menbart02}) have
recently explored the
   interest of this survey in order to cross-correlate the foreground
 galaxy distribution with the quasar population.  The expected
magnification 
  bias generated by dark matter associated with foreground 
  structures as mapped by galaxies  depends on $\Omega_m$ 
  and the biasing $\sigma_8$.   
    In principle magnification bias in
the 
SDSS quasar sample can provide similar constrains as cosmic
shear.
\section*{Aknowledgements}
 We thank M. Bartelmann, K. Benabed, D. Bond, 
T. Hamana, H. Hoekstra, B. M\'enard,
  S. Prunet and P. Schneider  for useful
   discussions.  
   This work was supported by the TMR Network
``Gravitational
 Lensing: New Constraints on
Cosmology and the Distribution of Dark Matter'' of the EC under contract
No. ERBFMRX-CT97-0172. YM thanks the organizers of the meeting for
  financial support.


\end{document}